\documentclass[11pt]{article}
\newcommand{\beq}{\begin{eqnarray}}% can be used as {equation} or {eqnarray}
\newcommand{\eeq}{\end{eqnarray}}
\newcommand{\drawsquare}[2]{\hbox{%
\rule{#2pt}{#1pt}\hskip-#2pt%  left vertical
\rule{#1pt}{#2pt}\hskip-#1pt%  lower horizontal
\rule[#1pt]{#1pt}{#2pt}}\rule[#1pt]{#2pt}{#2pt}\hskip-#2pt%  upper horizontal
\rule{#2pt}{#1pt}}% right vertical

\newcommand{\Yfund}{\raisebox{-.5pt}{\drawsquare{6.5}{0.4}}}%  fund
\newcommand{\Ysymm}{\raisebox{-.5pt}{\drawsquare{6.5}{0.4}}\hskip-0.4pt%
        \raisebox{-.5pt}{\drawsquare{6.5}{0.4}}}%  symmetric second rank
\newcommand{\Yasymm}{\raisebox{-3.5pt}{\drawsquare{6.5}{0.4}}\hskip-6.9pt%
        \raisebox{3pt}{\drawsquare{6.5}{0.4}}}%  antisymmetric second rank

\newcommand{\Yoneoone}{\raisebox{-3.5pt}{\drawsquare{6.5}{0.4}}\hskip-6.9pt%
        \raisebox{3pt}{\drawsquare{6.5}{0.4}}\hskip-6.9pt%
        \raisebox{9.5pt}{\drawsquare{6.5}{0.4}}\hskip-0.4pt%
        \raisebox{9.5pt}{\drawsquare{6.5}{0.4}}}%
\newcommand{\jref}[4]{{\it #1} {\bf #2}, #3 (#4)}

\newcommand{\NPB}[3]{\jref{Nucl.\ Phys.}{B#1}{#2}{#3}}

\newcommand{\PLB}[3]{\jref{Phys.\ Lett.}{#1B}{#2}{#3}}

\newcommand{\PRD}[3]{\jref{Phys.\ Rev.}{D#1}{#2}{#3}}

\newcommand{\PRL}[3]{\jref{Phys.\ Rev.\ Lett.}{#1}{#2}{#3}}

\makeatletter
\def\vereq#1#2{\lower3pt\vbox{\baselineskip1.5pt \lineskip1.5pt
\ialign{$\m@th#1\hfill##\hfil$\crcr#2\crcr\sim\crcr}}}
\makeatother

\begin{document}
\begin{center}
\today     \hfill    LBNL-41794 \\
%\hfill UCB-PTH-98/??  \\
\hfill hep-th/9805053\\
\vspace*{1cm}
\begin{large}
\textbf{'t HOOFT ANOMALY MATCHING FOR DISCRETE SYMMETRIES}\footnote{Talk
presented by Csaba Cs\'aki at  
the XXXIIIrd Rencontres de Moriond, Electroweak Interactions and 
Unified Theories, Les Arcs, Savoie, France, March 14-21, 1998.} \\
\end{large}
\vspace*{22pt}
{\bf Csaba Cs\'aki\footnote{Research fellow, Miller Institute for 
Basic Research in Science.} and Hitoshi Murayama\footnote{Alfred P. Sloan 
Foundation fellow.}} \\
\textit{Theoretical Physics Group\\
     Ernest Orlando Lawrence Berkeley National Laboratory\\
and \\
Department of Physics, University of California \\
 Berkeley, California 94720 }
\end{center}

\vspace*{1cm}

\begin{abstract}

We show how to extend the 't Hooft anomaly matching conditions to discrete 
symmetries. We check these discrete anomaly matching conditions on several 
proposed low-energy spectra of certain strongly interacting gauge theories. 
The excluded examples include the proposed chirally symmetric vacuum of pure 
$N=1$ supersymmetric Yang-Mills theories, certain non-supersymmetric confining
 theories and some self-dual $N=1$ supersymmetric theories based on 
exceptional groups.
\end{abstract}

\baselineskip 15pt

\section{'t Hooft Anomaly Matching for Continuous Global Symmetries}
\setcounter{equation}{0}
\setcounter{footnote}{0}
't Hooft anomaly matching~\cite{tHooft} is a powerful tool to constrain the 
massless fermionic bound-state spectrum. Finding the massless spectrum is 
very important, since it is the first step towards establishing an effective 
low-energy Lagrangian (an analog of the chiral Lagrangian for QCD) of a given 
strongly interacting theory. 't Hooft was arguing that the global symmetries 
can be used to severely restrict the massless fermion spectrum. Below we 
briefly summarize 't Hooft's original argument. Assume we have a strongly 
interacting gauge theory based on the gauge group $G_c$, and that the theory 
has in addition  a $G_F$ flavor symmetry. In order for the theory to be 
consistent, the gauge anomalies $G_c^3$ have to cancel. In order for the 
$G_F$ to be an unbroken 
global symmetry, the mixed $G_c^2G_F$ anomalies have to vanish as well. 
However,
a priori there is no reason for the $G_F^3$ anomalies, the anomalies 
calculated solely with respect to the global symmetries themselves, to 
vanish. It turns out that these $G_F^3$ anomalies, instead of being vanishing,
 will put a non-trivial constraint on the massless spectrum of the theory. 

To see this, introduce {\it spectator} fields, which {\it do not} transform 
under the strong gauge group $G_c$, only under the flavor symmetry $G_F$, 
such that all $G_F^3$ anomalies vanish. In this enlarged theory we can weakly
 gauge the $G_F$ flavor symmetry, and consider the low-energy limit of this 
modified theory. At low energies, the $G_c$ gauge group will confine 
the original degrees of freedom into bound states. However, since we can take 
the gauge coupling of $G_F$ to be arbitrarily small, we expect $G_F$ to be 
still weak at low energies. Thus the low-energy effective theory should be a 
weakly interacting $G_F$ gauge theory of the composite $G_c$ bound states.
In order for this effective theory to be consistent, the $G_F^3$ anomalies
still have to be vanishing in the effective theory. Now let us compare the
extended theory to our original one. We notice, that since the 
spectators do not transform under $G_c$, they do not participate 
in forming the bound states. Thus they are included both into the high-energy
and low-energy theories as elementary fields, therefore their contribution
to the $G_F^3$ anomalies is identical in the low-energy and in the
high-energy theories. Thus the remaining degrees of freedom also have to have 
matching $G_F^3$ anomalies: {\it the global anomalies of the 
elementary degrees of freedom have to match those of the massless bound states,
if $G_F$ is not spontaneously broken}. This statement is 't Hooft anomaly 
matching. It is a set of necessary conditions which the correct low-energy
spectrum has to satisfy, and which played a central role in establishing
exact results in $N=1$ supersymmetric gauge theories~\cite{Seiberg}.
Explicitly, the expressions for the anomalies which have to be matched
 for a $G_c$ gauge theory with global symmetry
$G_{F}=G_1\times G_2 \times \ldots \times U(1)_1\times 
U(1)_2 \times \ldots$ are: $ G_i^3:  \sum_R A^i_R $, $ G_i^2U(1)_j:  \sum_R
 \mu_R^iq_R^j$, $  U(1)_iU(1)_jU(1)_k:  \sum_R q_R^iq_R^jq_R^k $, 
$  U(1)({\rm gravity})^2:  \sum_R q_R^i $, 
where $A$ is the cubic anomaly coefficient defined by the relation
${\rm Tr}_R\, \{ T^a,T^b\} T^c=A_Rd^{abc}$ (the
$T$'s being the generators of the group $G_i$ in a given representation $R$),
$\mu_R$ is the Dynkin index ${\rm Tr}_R\, T^aT^b=\mu_R \delta^{ab}$,  
and $q^i$'s are the
$U(1)_i$ charges. The sum over $R$ denotes the summation over all 
representations
of fermions present in the high-energy or the low-energy descriptions.

\section{Discrete Anomaly Matching}
\setcounter{equation}{0}
\setcounter{footnote}{0}
Following the logic of the previous section, the following question arises 
naturally: can we extend the 't Hooft anomaly matching conditions to discrete 
global symmetries as well? We will show, that the answer is yes, however, 
these conditions are weaker than those for continuous symmetries. We consider 
only abelian $Z_N$ discrete symmetries. Since the $Z_N$ charges are defined
only mod $N$, the best we can hope for are matching conditions that
have to be satisfied mod $N$. We will see, that indeed, some of the discrete
anomaly matching conditions will be mod $N$, but some of them slightly
weaker. 

Consider first the case of the $G_F^2Z_N$ anomalies, where $G_F$ is a 
simple Lie group, and where $G_F$ and $Z_N$ are assumed to be global
symmetries of our gauge theory with gauge group $G_c$. As in 't Hooft's 
argument, we can include spectators which do not transform under the gauge
group, only under $G_F$ and $Z_N$, such that the $G_F^3$ and the $G_F^2Z_N$
anomalies vanish. Now we can weakly gauge the $G_F$ global symmetry.
Since the $G_F^2Z_N$ anomaly vanishes, that is $\sum_i \mu_i q_i=0\; 
 {\rm mod}\;
N$ where $\mu$ is the Dynkin index and $q$ is the $Z_N$ charge, 
the $Z_N$ discrete symmetry is unbroken even in the background of
$G_F$ instantons. Now we consider the low-energy effective
theory. Since the $Z_N$ is a good symmetry of the full extended theory,
the $G_F^2Z_N$ anomalies have to vanish in the low-energy effective 
theory as well, thus  $\sum_i \mu_i q_i$ still vanishes mod $N$. Since the
spectators do not transform under the strong gauge group $G_c$, their
contribution to  $\sum_i \mu_i q_i$ is identical in the high-energy and the
low-energy theories. Thus we conclude that the $G_F^2Z_N$ anomalies 
have to be matched mod $N$. Note, that in this argument, one never had to
promote the discrete symmetry to a continuous one,
contrary to the criticism of Ref.~\cite{KKS}. Therefore, the objection
raised in Ref.~\cite{KKS} has no basis.

Similarly, one can consider the $Z_N({\rm gravity})^2$ anomaly, which 
constrains the quantity $\sum_i q_i$. Considering correlators  in gravitational
instanton backgrounds, we conclude that $\sum_i q_i$ has to be matched 
mod $N/2$. The origin of the (weaker) mod $N/2$ matching is that 
every fermion has at least two zero modes in gravitational
instanton backgrounds. An alternative explanation for the possible 
additional $N/2$ contribution to this this anomaly is to note that 
a heavy Majorana fermion with $Z_N$ charge $N/2$ does not have a 
vanishing $Z_N({\rm gravity})^2$ anomaly, instead it exactly
contributes $N/2$. 

Thus we conclude, that the $G_F^2Z_N$ anomaly has to be matched mod $N$, and
the $Z_N({\rm gravity})^2$ anomaly mod $N/2$. We call these the Type I
anomalies. For the remaining (Type II) anomalies ($Z_N^3$, $U(1)^2Z_N$,
$U(1)Z_N^2$, $Z_N^2Z_M$) there is no similar argument in favor of anomaly
matching based on instantons. However, by promoting certain parameters
of the Lagrangian to background fields one can extend the discrete symmetry
to a continuous global $U(1)$ symmetry, from which one can argue that 
the Type II anomalies still have to be matched mod $N$ (for more details see
Ref.~\cite{discrete}). There are however some important subtleties which one 
has to consider for these Type II anomalies:

- Decoupling of Majorana fermions can weaken the $Z_N^3$ matching condition,
just like in the case of  the $Z_N({\rm gravity})^2$ anomaly

- One has to choose a normalization where all $U(1)$ charges are integers
in order to have a mod $N$ matching condition

- Existence of fractionally charged massive states can invalidate the
matching of Type II anomalies (but not that of Type I anomalies)

Thus, we conclude that the discrete anomalies have to be matched in the
following way: Type I:  $G^2Z_N$: $mN$,  $Z_N(\mbox{gravity})^{2}$:  $mN 
+\frac{m'}{2}N$; Type II:  $Z_N^3$: $mN + \frac{m'}{8}N^3$,
 $ U(1)^2 Z_N$: $mN$, $ U(1)_i U(1)_j Z_N$: $mN$,
 $ U(1)Z_N^2$:  $mN$, $ U(1)Z_NZ_M$:  $mK$, where after each anomaly we 
have indicated the possible difference.
Here  $m, m'$ 
  are integers, and $m'$ can be non-vanishing 
  only if $N,M$ are even.  $K$ is the GCD of $N$ and $M$. Type I
  anomaly matching constraints have to be satisfied regardless of
  the details of the massive spectrum. Type II anomalies have to be
  also matched except if there are fractionally charged
  massive states.

\section{An Example}
\setcounter{equation}{0}
\setcounter{footnote}{0}
We have checked, that all Seiberg dualities~\cite{Seiberg}, 
including Kutasov type dualities~\cite{Kutasov} (at least the ones we have 
checked from the long list of theories in~\cite{Kutasov}) satisfy the 
discrete anomaly matching conditions presented in the previous section, even 
though some of these matching conditions are very non-trivial. Here we
present only one simple example, which is based on an s-confining
$N=1$ supersymmetric 
theory~\cite{s-conf}. The theory together with the confining spectrum 
is given in the table below.

\[ \begin{array}{c|c|ccc}
& SO(7) & SU(6) & U(1)_R & Z_{12} \\ \hline
S & 8 & \Yfund & \frac{1}{6} & 1 \\ \hline \hline
S^2 & & \Ysymm & \frac{1}{3} & 2 \\
S^4 & & \overline{\Yasymm} & \frac{2}{3} & 4 \end{array} 
\nonumber \]
$SO(7)$ is the gauge group and $ SU(6)\times U(1)_R\times Z_{12}$ are the
global symmetries.
The anomaly matching conditions are:
\[
\begin{array}{c|c|c}
& \mbox{UV} & \mbox{IR} \\
\hline
 SU(6)^2Z_{12}  &  8   &
 2\times 8+4\times 4=8+2\times 12  \\
 Z_{12}(\mbox{gravity})^{2}  &  48  & 
  2\times 12+4\times 15 = 8\times 12 + 6 \\
 Z_{12}^3 &  48  & 
 2^{3}\times 21+4^{3}\times 15=94\times 12\\
  U(1)_R^2 Z_{12} &  1200  &
 76\times 12 \\
  U(1)_R Z_{12}^2 &  -5\times 8\times 6  &  -68\times 12 
\end{array} \nonumber \]
where the contributions to the first three anomalies in the magnetic theory 
are quoted in the order $S^{2}$, $S^{4}$.  The $U(1)_{R}$ 
charges are multiplied by a factor of 6 to make all the charges 
integers.

All anomalies match mod $12$ except the $Z_{12}(\mbox{gravity})^{2}$
anomaly, which is matched mod $6$, and signals the presence of
massive Majorana fermions with charge $6$.  But we do not see the 
corresponding contribution to $Z_{12}^{3}$ anomaly because $12^{3}/8 = 
216$ is a multiple of 12.  

\section{Excluded Examples}
\setcounter{equation}{0}
\setcounter{footnote}{0}
Kovner and Shifman argued recently~\cite{Shifman}, that there might be an 
additional, chirally symmetric phase of $N=1$ pure Yang-Mills theory,
with vanishing gaugino condensate. In this vacuum, the $Z_{2N}$ 
(for the case of $SU(N)$ gauge groups) discrete
$R$-symmetry  is not broken, therefore
the discrete anomalies have to be matched by the massless fields of the 
low-energy effective theory. The most natural candidate for the massless
field is $\Phi=(W_{\alpha}W^{\alpha})^{\frac{1}{3}}$, since this is the
basic variable of the Veneziano-Yankielowicz Lagrangian~\cite{VY}, 
based on which
Kovner and Shifman concluded that there might be an additional vacuum. 
In this case, the $R$-charge of the fermionic component of $\Phi$ is
$-\frac{1}{3}$, which signals the fractionalization of the $Z_{2N}$ charges. 
Therefore, it is convenient to rescale the discrete
charges such that the gaugino of the high-energy theory has charge
$3$, and check the anomaly matching conditions for the resulting 
$Z_{6N}$ symmetry. The discrete anomalies for $SU(N)$ are:
\[
\begin{array}{c|c|c}
& \mbox{UV} & \mbox{IR} \\ \hline
Z_{6N} ({\rm gravity})^2 & 3(N^2-1) & -1 \\
Z_{6N}^3 & 27(N^2-1) & -1 
\end{array} \nonumber \]
The difference in the $Z_{6N} ({\rm gravity})^2$ anomalies of the UV and the IR
descriptions is 2 mod $3N$, which means that the discrete anomalies can
not be matched for any value of $N$.  Recall that the $Z_{6N} ({\rm
  gravity})^2$ anomaly is Type I and 
must be matched irrespective of charge
fractionalization.  Therefore, this low-energy description of the pure
$SU(N)$ YM theories is excluded. One can show in an analogous way, that the
Kovner-Shifman vacua are excluded by discrete anomaly matching for the
other simple groups as well. 

However, this does not completely exclude
the idea of a chirally symmetric phase of $N=1$ pure
Yang-Mills theories. It excludes only a specific realization of it
described above. One
could, for example, try to match anomalies with the operator $S= 
W_{\alpha}W^{\alpha}$ instead of $\Phi$. 
Here no charge fractionalization occurs, and hence anomalies should be matched
mod $2N$.
The anomalies for $SU(N)$ are
\[ \begin{array}{c|c|c}
& \mbox{UV} & \mbox{IR} \\ \hline
Z_{2N} ({\rm gravity})^2 &  N^2-1 & 1 \\
Z_{2N}^3 & N^2-1 & 1
\end{array} \nonumber \]
The differences in the anomalies are both $N^2-2$, which is divisible by $N$
only for $N=1,2$. Performing a similar analysis we find that the field
$S$ matches the discrete anomalies for $SO(N)$ only if $N$ is odd, while
it matches always for $Sp(2N)$.  None of the discrete anomalies for the
exceptional groups are matched by $S$.  Even
though anomalies are matched for some special cases by $S$,
generically it does not match the discrete anomalies and therefore 
we conclude that it is not a likely candidate for a low-energy solution. 

In addition to the Kovner-Shifman vacuum, the excluded examples include 
several non-supersymmetric theories which were conjectured to be
confining in the early 80's~\cite{Albright}. An example of such a theory 
with the conjectured low-energy spectrum is given in the table below.

\[ \begin{array}{c|c|ccc}
& SU(4)& SU(2) & U(1) & Z_{12} \\ \hline
A & \Yasymm & \Yfund & 2 & 1 \\
X & \Yoneoone & 1 & -1 & 1 \\ \hline \hline
(A^2X) & 1 & \Ysymm & 3 & 3 \end{array} \nonumber 
\]
All the continuous global anomalies ($SU(2)^2U(1)$,
$U(1)({\rm gravity})^2$ and $U(1)^3$) are matched between the
high-energy and the
confining spectrum. The discrete anomalies are:
\[
\begin{array}{c|c|c}
& \mbox{UV} & \mbox{IR} \\ \hline
SU(2)^2Z_{12} & 6 & 12 \\
Z_{12} ({\rm gravity})^2 & 27 & 9\\
Z_{12}^3 & 27 & 81 \\
U(1)^2Z_{12} &  63 & 81 \\
U(1) Z_{12}^2 & 9 & 81 
\end{array} 
\nonumber \]
The $ U(1)^2Z_{12}$ anomaly matching is satisfied mod $12$ and
the $ Z_{12} ({\rm gravity})^2$ anomaly matching is 
satisfied mod $6$. However, while the $SU(2)^2Z_{12}$, the $
U(1)^2Z_{12}$ and the $Z_{12}^3$ 
anomalies must match mod $12$, they match only
mod $6$,  and hence the discrete anomaly matching conditions are
violated. In the absence of any dynamical explanation of spontaneous
breaking of $Z_{12}$, and since $SU(2)^2Z_{12}$ is a Type I anomaly,
 one has to consider this model excluded  based on
discrete anomaly matching. 

Similarly, certain $N=1$ supersymmetric dualities based on exceptional 
groups can be excluded as well using the discrete anomaly matching conditions
(see \cite{discrete} for details).

Finally, we comment on an interesting example, where continuous 
anomaly matching can lead to misleading conclusions~\cite{BCI}.
The theory is $N=1$ $SO(N)$ with a symmetric tensor.  
All continuous anomalies are matched by the set of independent gauge
invariant operators, and the Type I discrete anomalies match as well
in this example.
However, the Type
II conditions are not satisfied.  As explained before, the failure of
the matching of Type II anomalies does 
not automatically exclude a given low-energy spectrum due to the 
possibility of charge fractionalization.
However, we have to emphasize that all established theories satisfy
the Type II conditions as well, and that the charge fractionalization
of the heavy states is quite unlikely.  Thus this raises the suspicion
that this theory is not confining.  Indeed, it was noted
in \cite{BCI} that there are several reasons to
believe that the theory is not confining at the origin.  This is a
good example where  the failure of Type II discrete anomaly matching is
the first sign of the incorrect guess on the low-energy dynamics.

\section{Conclusions}
We have shown how to extend the 't Hooft anomaly matching conditions to
discrete global symmetries. There are two types of discrete anomalies. 
Type I anomaly matching conditions ($G_F^2Z_N$ and $Z_N({\rm gravity})^2$) 
have to be satisfied
regardless of assumptions on the massive bound states. Type II
constraints have to be satisfied except if there are fractionally
charged massive states. We have tested several conjectured low-energy
solutions using discrete anomaly matching. The excluded examples are:
the chirally symmetric phase of $N=1$ pure Yang-Mills theories,
certain non-supersymmetric theories conjectured to be confining, and 
$N=1$ supersymmetric self-dualities based on exceptional groups.

\section*{Acknowledgements}
This work was 
supported in part by the U.S. Department of Energy under Contracts 
DE-AC03-76SF00098 and in part by the National Science Foundation under 
grant PHY-95-14797. C.C. is a research fellow of the Miller Institute
for Basic Research in Science. H.M. is an Alfred P. Sloan Foundation
fellow.

\end{document}